\documentclass{article}
\usepackage[T1]{fontenc}
\usepackage{spconf,amsmath,graphicx,hyperref}
\usepackage{booktabs}
\usepackage{amsmath}
\usepackage{algorithm}
\usepackage{algpseudocode}
\usepackage{placeins}
\usepackage{etoolbox}

\title{Hiding Tool Latency in On-Device Cascaded Voice Agent\\ through Speculative Execution}
\name{
\textit{Kyudan Jung$^{1,2}$\thanks{This publication was created by Kyudan Jung while an intern at Qualcomm Technologies, Inc. (QTI) and currently attends Korea Advanced Institute of Sceience and Technology (KAIST).}, Hyunsin Park$^{1}$, Yoonhyung Lee$^{1}$, Jinhwan Park$^{1}$,}\\
\textit{Jinhyeok Yang$^{1}$, KiHyun Nam$^{1}$, Jaegul Choo$^{2}$, Jinkyu Lee$^{1}$}}
\address{$^{1}$Qualcomm AI Research, $^{2}$KAIST AI}

\begin{document}
\maketitle
\begin{abstract}

Tool-augmented speech assistants typically serialize automatic speech recognition, large language model inference, and external tool execution.
As a result, tool latency is incurred only after the user has finished speaking and the LLM has identified the required tool calls.
We present speculative tool execution for on-device cascaded voice agents, which predicts tool requests from partial ASR hypotheses and initiates tool execution while speech is still being received, thereby reducing end-to-end response latency.
Our approach introduces a Predictor module that anticipates tool calls during speech recognition, executes them speculatively, and caches the results.
The cached outputs are then injected into the LLM prompt, enabling faster responses.
Additionally, to mitigate errors caused by user self-corrections during speech, we employ a rule-based validation mechanism that selectively injects only valid cached results.
As a final safeguard, the LLM retains the ability to issue tool calls directly, ensuring that the latency of our framework is upper-bounded by the baseline serial execution pipeline in the worst case.
We evaluate our method using live measurements from a fully implemented Android voice assistant. Our approach reduces the median time-to-first-audio from 5.79\,s to 4.60\,s and decreases the standard deviation from 3.49\,s to 2.81\,s, resulting in more predictable response latency.\end{abstract}


\section{Introduction}
\label{sec:intro}
Cascaded speech assistants connect automatic speech recognition (ASR), a large language model (LLM), and text-to-speech (TTS) in sequence~\cite{he2019streaming,li2017alexa}, allowing explicit text editing between modules and thus providing greater reliability than end-to-end systems.
These assistants can invoke tools for a wide range of tasks, from adding calendar events to searching the web~\cite{qin2024toolllm,li2023apibank,huang2024metatool,schick2023toolformer,yao2023react}.
Unlike end-to-end speech systems that can initiate tool calls during streaming generation~\cite{arora2025streamraginstantaccurate}, conventional cascaded assistants must wait for the user to finish speaking, finalize the transcript, and obtain a tool request from the LLM before invoking external tools.
While the latency of simple local tools may be negligible, remote operations such as time-consuming web search and can introduce substantial delays that are directly perceived by users.
Reducing this latency is therefore important for responsive voice interactions on mobile devices~\cite{kim2024llmcompiler,gim2024async}.

Streaming speech provides an opportunity to reduce this delay because user intent can often become apparent before the utterance is complete~\cite{he2019streaming, gulati2020conformer}.
For example, when a request can be predicted from a partial transcript, its execution can begin while the user is still speaking, overlapping tool latency with the remaining ASR time. This perspective treats streaming speech not only as an input modality but also as an opportunity for anticipatory computation~\cite{zhang2023speechgpt,defossez2024moshi,navercloudhyperclovaxteam2026hyperclovax8bomni}.
\begin{figure}[t]
\centering
\includegraphics[width=0.5\textwidth]{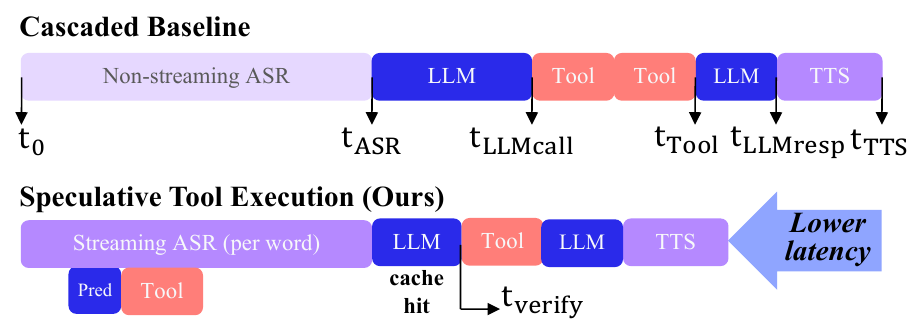}
\vspace{-20pt}
\caption{Timing block diagram of speculative tool calling. Unlike the cascaded baseline, our method hides part or all of the tool-calling latency during ASR. The figure illustrates a case where one of two required tool calls results in a cache hit. }
\vspace{-10pt}
\label{fig:overall}
\end{figure}
Based on this observation, we introduce a rule-based predictor that continuously observes partial ASR words, predicts likely long-latency tool requests, and launches them before the LLM explicitly requests the corresponding tools.
The resulting outputs are stored in a speculative result cache and can later be reused by the LLM execution pipeline~\cite{arora2025streamraginstantaccurate,chen2026mistmultimodalinteractivespeechbased,laskar2026textvoicereproducibleverifiable}.
We then inject the retrieved result directly into the LLM prompt before response generation, allowing the model to use information obtained while the user is still speaking. The LLM retains its standard tool-calling path as a fallback when no suitable speculative result is available.
\begin{figure*}
    \centering
    \includegraphics[width=1\linewidth]{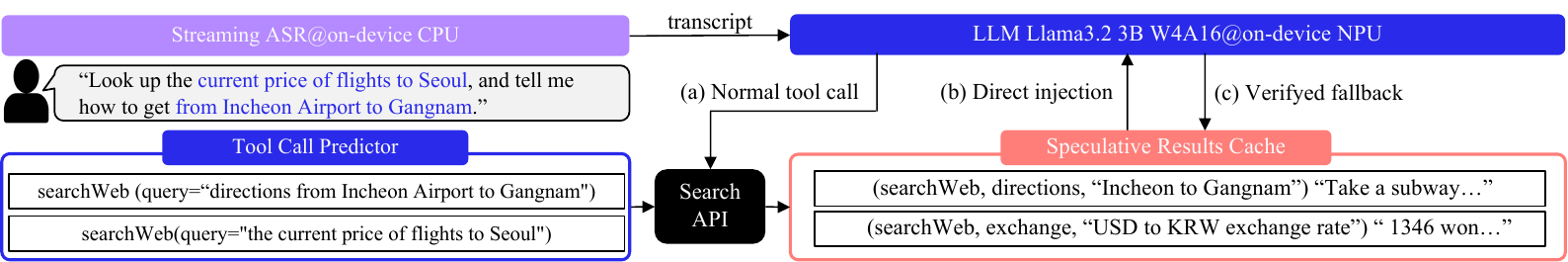}
    \vspace{-20pt}
    \caption{Overview of our methodology. During streaming ASR processing, the Predictor anticipates potential tool calls and retrieves relevant information via a Search API. The retrieved results are stored in the Speculative Results Cache and directly injected into the LLM through path (b). When the LLM later issues the corresponding tool call, the system first checks the cached results through path (c) for verification. As a fallback, the standard tool-calling workflow is executed through path (a), corresponding to the conventional LLM tool invocation process. In our setup, the ASR model runs on the CPU, while the W4A16 quantized LLM runs on the NPU.}
    \label{fig:algorithm_method}
\end{figure*}

 We evaluated our method in an on-device to analyze its performance under resource-constrained conditions. The median waiting time from the end of the user’s speech to the system’s response decreased from 5.79 seconds to 4.60 seconds. In particular, fewer high-latency outliers led to reduced variance and more predictable response times, while the tool-calling F1 score increased from 44.8\% to 60.6\%. The ablation study shows that direct cache injection is the main source of the latency improvement but also accuracy, while verified fallback increases cache reuse and reduces wasted speculative calls. The system remains robust to hard-negative and adversarial requests, and the neural predictor analysis suggests that the lightweight rule-based predictor provides a practical balance between early prediction and on-device inference cost.

\section{Methodology}
\label{sec:method}
We first formulate the problem and define the relevant timestamps. We then describe the proposed algorithm, including the execution timing of each component.

\subsection{Problem Formulation}
Based on Figure~\ref{fig:overall} baseline, we define the timestamp first. Let $t_0$ denote the start time of the user's utterance. Let $t_{\mathrm{ASR}}$ be the time at which the ASR system finalizes the transcript. After transcript finalization, the first LLM inference determines whether a tool should be invoked and, if so, which tool to call. We denote the completion time of this stage as $t_{\mathrm{LLM\text{-}call}}$. The selected tool is then executed, and its completion time is denoted by $t_{\mathrm{Tool}}$. Subsequently, a second LLM inference generates the textual response to be provided to the user, whose completion time is denoted by $t_{\mathrm{LLM\text{-}resp}}$ which means LLM response. Finally, the generated response is passed to the text-to-speech (TTS) system, and $t_{\mathrm{TTS}}$ denotes the completion time of speech synthesis. The timeline can therefore be expressed as \( t_0 < t_{\mathrm{ASR}} < t_{\text{LLMcall}} < t_{\mathrm{Tool}} < t_{\text{LLMresp}} < t_{\mathrm{TTS}}=t_{\text{End}}. \) We will hide time $t_{\mathrm{Tool}}-t_{\text{ASR}}$ with ongoing ASR processing, thereby reducing the total time $t_{\mathrm{TTS}}-t_0$.

\subsection{Predicting Which Tool to Call}
To determine which tool to call while ASR is operating, we introduce a predictor module which checks each partial ASR hypothesis during streaming. It runs after a minimum number of words $N_{\min}$ is observed, after every $\Delta N$ additional words, or when a utterance ended pause longer than $\tau_{\mathrm{pause}}$ is detected. Based on the real case distribution, we use $N_{\min}=6$, $\Delta N=5$, and $\tau_{\mathrm{pause}}=300$ ms.

The predictor first divides the current partial ASR hypothesis into non-overlapping clauses. The accumulated partial hypothesis available at each prediction step is segmented using sentence-boundary punctuation and coordinating expressions such as \emph{and}, \emph{but}, \emph{also}, \emph{then}, and \emph{while}. Each resulting clause is examined independently. As the ASR hypothesis grows, the predictor repeats this procedure at the predefined trigger points.
For example, the partial hypothesis ``\textit{Check the weather in Boston, then find nearby restaurants, and tell me the yen exchange rate}'' is divided into three clauses: ``\textit{Check the weather in Boston,}'' ``\textit{find nearby restaurants,}'' and ``\textit{tell me the yen exchange rate.}'' These clauses may produce the candidate tuples (\text{searchWeb}, \text{weather}, \text{``weather in Boston''}), (\text{searchWeb}, \text{place}, \text{``nearby restaurants''}), and (\text{searchWeb}, \text{exchange}, \text{``JPY--USD exchange rate''}), respectively. The topic field is drawn from a fixed set of 12 coarse-grained information-seeking categories. Thus, the clause-level analysis allows multiple independent information-seeking intents in a single utterance to be dispatched in parallel. Our Prdictor is motivated by the idea of incremental spoken-language understanding \cite{jorgensen2007clause}.

\subsection{Direct Injection and Cache Matching}
At $t_{\mathrm{ASR}}$, the finalized transcript and completed speculative results are directly inserted into the LLM prompt before decoding begins, as shown in Figure~\ref{fig:algorithm_method}(b). The LLM first generates an output $y$ conditioned on both the transcript and the cached results. Nevertheless, in our evaluation set, the LLM still issued tool calls in 13\% of cases, and due to the limitations of the small quantized language model, 7\% of cases redundantly requested tools for information that had already been provided. If the LLM nevertheless issues a tool call, the system checks whether a matching result is already available in the speculative cache at time $t_{\text{verify}}$ in Figure~\ref{fig:overall}, thereby preventing redundant retrieval requests, as shown in Figure~\ref{fig:algorithm_method}(c). When a matching entry exists, the cached result is reused and returned to the LLM. Otherwise, the requested tool is executed through the normal path, as in Figure~\ref{fig:algorithm_method}(a). This design enables the immediate use of results obtained during the user's speech while retaining LLM-directed tool calling and normal tool execution as reliable fallback paths. An ablation study was conducted to evaluate the contributions of direct injection and cache-verification fallback.

\begin{table}[t]
\centering
\vspace{-6pt}
\caption{Main results. Cascaded denotes the non-speculative baseline. TTFA is measured from the end of speech input to the first TTS audio output. FT denotes the time from ASR completion to the \textbf{F}irst \textbf{T}ool call.}
\label{tab:main_results}
\resizebox{\columnwidth}{!}{%
\begin{tabular}{lrrrrr}
\toprule
System & p50 & p99 & FT p99 & Mean $\pm$ Std & F1  \\
 & (s) & (s) & (s) & (s) & (\%)  \\
\midrule
Cascaded & 5.79 & 19.76 & 22.44 & 6.69 $\pm$ 3.49 & 44.8  \\
\textbf{Speculative (Ours)} & \textbf{4.60} & \textbf{15.10} & \textbf{11.71}& \textbf{5.99 $\pm$ 2.81} & \textbf{60.6}  \\
\bottomrule
\end{tabular}%
}
\end{table}
\section{Experiments}
\label{sec:experiments}

\subsection{Experimental Setup}
We evaluate on a Samsung Galaxy S26 Ultra (SM-S948N, Snapdragon Elite Gen 5) with the hardware organization. Model inference uses a streaming Conformer ASR model~\cite{gulati2020conformer} and a TTS model~\cite{ren2020fastspeech} running on the CPU, while a quantized Llama-3.2-3B QAIRT LLM bundle~\cite{grattafiori2024llama3herdmodels, qualcomm_llama32_3b_instruct} is executed through the Qualcomm QNN stack on the NPU. CPU-side components handle audio I/O, the rule-based speculative predictor, cache management, and HTTP dispatch to the EXA remote web-search API~\cite{exa_search_api}.

\subsection{Evaluation Dataset}
There is no public dataset designed to evaluate tool use in an Android voice assistant with tasks such as web search including android calendar events, and alarms. We therefore construct an evaluation set that covers these tasks and includes both normal requests and challenging cases.
The dataset contains 100 manually written English utterances covering single- and multi-search requests, calendar and alarm requests, and also requests that require no tool. Each utterance is annotated with the expected tools and arguments. We additionally include 22 hard-negative examples that contain search-related expressions but do not require a search, and 24 adversarial examples that include corrections, retractions, late intents, or topic changes. This gives a total of 146 evaluation samples.
For reproducible evaluation, we synthesize each utterance using Gemini-2.5-Flash-Preview-TTS~\cite{google_gemini_api_pricing} and stream the generated audio into the device in real time. The full evaluation dataset contains approximately 1.5 hours of audio, therefore a complete run takes approximately 3 hours including TTS generation and live network calls.

\subsection{Metrics}
Our primary latency metric is time-to-first-audio (TTFA). We measure TTFA from the end of the input speech to the first TTS audio output.
We evaluate search performance separately from general tool planning. A read request is counted as successful when the required \texttt{searchWeb} tool result is used in the final answer.
We report precision, recall, F1, and exact accuracy for tool requests. False read measures the fraction of turns that do not require a search tool but still use a search result. Wasted calls are speculative tool calls that results are not required to make a response. Cache-ready rate measures the fraction of required tool calls for which a matching speculative result is already available before the tool result is needed.

\begin{table}[t]
\centering
\vspace{-6pt}
\caption{Ablation study. Components are added incrementally to the cascaded baseline. Ready measures cache availability when a search result is needed, and Wasted counts speculative tool calls whose outputs are never consumed.}
\label{tab:ablation}
\resizebox{\columnwidth}{!}{%
\begin{tabular}{lrrrrrr}
\toprule
Variant & p50 & p95 & Mean $\pm$ Std &  F1 & Ready & Wasted \\
 & (s) & (s) & (s) & (\%) & (\%) & calls \\
\midrule
cascaded & 5.79 & 11.92 & 6.69 $\pm$ 3.49 & 44.8 & 0.0 & 0 \\
speculative prefetch only & 6.03 & 11.19 & 6.42 $\pm$ 2.69 & 44.8 & 0.0 & 47 \\
verified fallback & 6.05 & 11.35 & 6.53 $\pm$ 2.97 & 43.5 & 10.2 & 38 \\
direct cache injection & 5.72 & \textbf{11.05} & 6.17 $\pm$ 2.68 & \textbf{63.0} & 46.6 & 6 \\
 \textbf{direct injection + verified fallback} & \textbf{4.60} & 11.39 & \textbf{5.99 $\pm$ 2.81} & 60.6 & \textbf{47.7} & \textbf{5} \\
\bottomrule
\end{tabular}%
}
\vspace{-12pt}
\end{table}
\vspace{-4pt}
\section{Results}
\vspace{-8pt}
\subsection{Main Results}
\label{main_results}
Table~\ref{tab:main_results} shows that our speculative system achieves faster and more consistent response times than the cascaded baseline. Its primary advantage comes from overlapping remote tool execution with the user's speech, thereby hiding a substantial portion of the search latency. For example, for a 15-second user query requiring approximately 10 seconds of search-tool execution, our method reduced the time to first audio (TTFA) from 17 seconds to 8 seconds. By mitigating such latency outliers in p99, speculative execution improves both the responsiveness and the perceived reliability of the voice agent.

Our method also improves tool-selection accuracy, leading to a higher F1 score and fewer unnecessary speculative tool calls. We hypothesize that this improvement arises from providing the small 3B language model with additional task-relevant context during tool selection. This observation is consistent with prior work showing that smaller language models can benefit substantially from informative prompts and in-context examples~\cite{gao2021making}. Thus, the proposed system improves not only the timing of tool execution but also the efficiency and accuracy with which tools are selected.

Speculative execution also makes response timing less sensitive. In the cascaded baseline, remote tool calls begin only after ASR and LLM-based tool selection have completed, causing end-to-end latency to depend heavily on potentially unstable network delays. In contrast, our method shifts a substantial portion of this variable latency into the user's speaking time. Consequently, the latency remaining after the user finishes speaking is dominated by local model inference, which is comparatively stable. This results in a faster and more predictable interaction experience without increasing the final-answer failure rate.

\begin{table}[t]
\centering

\caption{Robustness diagnostics, with key outcomes in bold. N/A denotes an inapplicable metric.}
\label{tab:robustness}
\resizebox{\columnwidth}{!}{%
\begin{tabular}{lrrrrrr}
\toprule
Set & $n$ & p50 &  F1 & False read & Spec calls & Wasted \\
 & & (s) & (\%) & (\%) & & calls \\
\midrule
Hard negatives & 22 & 6.24 & N/A & \textbf{9.1} & 2 & 2 \\
Adversarial & 24 & 8.40 & \textbf{90.2} & 12.3 & 21 & \textbf{21} \\
\bottomrule
\end{tabular}%
}
\vspace{-15pt}
\end{table}

\vspace{-10pt}
\subsection{Ablation Study}

Table~\ref{tab:ablation} summarizes the contribution of each component. The ``Speculative prefetch only'' variant runs speculative tool calls concurrently with on-device ASR while retaining the remaining baseline pipeline. Although its median latency increases slightly, its mean and p95 latency remain comparable to or lower than those of the cascaded baseline. This suggests that concurrent ASR and speculative prefetching introduce no noticeable latency penalty.

The ``Verified fallback'' variant allows the system to consult the cache only after the LLM has selected a tool. Its latency and F1 score remain close to those of the baseline, indicating that cache reuse alone provides limited benefit when the full tool-selection process still occurs after ASR. We attribute this result to the fact that the LLM must complete tool selection before the system can determine whether a matching result is available, leaving most of the tool-selection latency exposed and introducing an additional cache lookup.

Directly injecting prefetched results into the prompt provides a larger improvement, reducing mean latency from 6.69 to 6.17 seconds and increasing F1 score from 44.8\% to 63.0\%. We attribute this improvement to the in-context learning effect discussed in Section~\ref{main_results}. Our final method adds a verified fallback to this design and achieves the lowest median and mean latency, the highest Ready rate, and the fewest wasted calls. Although its p95 latency and F1 are slightly worse than those of direct cache injection alone, it provides the best overall balance between response speed, cache utilization, and speculative-call efficiency. These results indicate that direct cache injection is primarily responsible for the improvement in tool-selection accuracy, while verified fallback improves the effective reuse of speculative results.

\subsection{Hard Negatives and Adversarial Cases} Table~\ref{tab:robustness} evaluates robustness under two challenging conditions. Hard negatives contain search-related keywords but do not actually require a search action, whereas adversarial examples include user corrections, retractions, and topic changes that alter the intended tool request during speech. The system achieves a low false read rate of 9.1\% on hard negatives, indicating that it rarely triggers unnecessary searches based on keywords alone. F1 score is not reported for this set because no ground-truth search requests exist. On adversarial examples, the system maintains a F1 score of 90.2\% with a false read rate of 12.3\%, showing that speculative results are usually discarded correctly when the user's intent changes. The 21 wasted calls reflect the additional network cost of speculative execution, but because speculation is restricted to read-only tools, such errors do not revise alarms, calendars, or other user data.

\begin{table}[t]
\centering
\caption{Offline router accuracy reference. Fire position is the mean fraction of transcript words observed before the first prediction. Bold is the best value per column.}
\label{tab:classifier_ablation}
\resizebox{\columnwidth}{!}{%
\begin{tabular}{lrrrr}
\toprule
Router input &  F1 $\uparrow$ & Exact count $\uparrow$ & False read $\downarrow$ & Fire position $\downarrow$ \\
 & (\%) & (\%) & (\%) & (\%) \\
\midrule
Rule word prefix & 37.5 & 49.0 & 22.2 & 50.6 \\
Rule clause prefix & 36.5 & 49.0 & 22.2 & 70.6 \\
Rule full utterance & 39.4 & 50.0 & 22.2 & 100.0 \\
LLM prefix 50\% & 40.0 & 67.0 & \textbf{8.9} & \textbf{49.8} \\
LLM prefix 75\% & 51.9 & 68.0 & 22.2 & 74.8 \\
LLM full utterance & \textbf{59.9} & \textbf{83.0} & 22.2 & 100.0 \\
\bottomrule
\end{tabular}%
}
\vspace{-10pt}
\end{table}

\subsection{Can an LLM based Predictor Perform Better?}
Table~\ref{tab:classifier_ablation} shows that a neural predictor can provide substantially higher routing accuracy than the proposed rule-based predictor. However, these results are obtained from an offline experiment using the on-device LLM as the same in experiments and do not account for its inference latency. In practical deployment, the LLM predictor should complete tool-routing decisions while the user is still speaking in order to create an overlap window for speculative execution. Given the runtime cost of on-device LLM inference, it is unlikely that the prediction can consistently finish before ASR finalization, which limits the opportunity for hiding tool latency. Running the LLM predictor earlier provides only a modest advantage over the lightweight rule-based predictor while introducing additional competition for on-device NPU resources. Therefore, our design favors a simple rule-based predictor, while delegating final semantic verification to the main LLM.

\vspace{-10pt}
\section{Conclusion}
\vspace{-10pt}

We presented a speculative tool execution method for an on-device voice agent that starts tool calls during streaming ASR. On a commercial Android device, our method reduced the latency and improved F1 accuracy. The results show that direct cache injection is the main source of the accuracy improvement, while verified fallback increases cache reuse and reduces wasted calls. Overall, our method makes voice-agent responses faster, more accurate, and more stable.

\section{ACKNOWLEDGMENTS}
The authors used Copilot for implement and refactor the experimental scripts used in Section 4. No text, figures, or experimental results were generated without author verification.

\bibliographystyle{IEEEbib}
\bibliography{refs}

\end{document}